# Characteristics of the Accelerated Electrons Moving along the Loop Derived from Cyclical Microwave Brightenings at the Footpoints

Keitarou Matsumoto[1] , Satoshi Masuda[1] , and Takafumi Kaneko[2]

[1] Institute for Space-Earth Environmental Research, Nagoya University, Furo-cho, Chikusa-ku, Nagoya, Aichi, 464-8601, Japan;
keitaromatsumoto@isee.nagoya-u.ac.jp
[2] Faculty of Education, Niigata University, 8050 Ikarashi 2-no-cho, Nishi-ku, Niigata, 950-2181, Japan

## Abstract

Many particles are accelerated during solar flares. To understand the acceleration and propagation processes of electrons, we require the pitch-angle distributions of the particles. The pitch angle of accelerated electrons has been estimated from the propagation velocity of a nonthermal microwave source archived in Nobeyama Radioheliograph data. We analyzed a flare event (an M-class flare on 2014 October 22) showing cyclical microwave brightenings at the two footpoint regions. Assuming that the brightenings were caused by the accelerated electrons, we approximated the velocity parallel to the magnetic field of the accelerated electrons as $\sim 7.7 \times 10^4$ and $9.0 \times 10^4$ km s$^{-1}$. The estimated pitch angle of the accelerated electrons is 69°–80° and the size of the loss cone at the footpoint (estimated from the magnetic field strength in the nonlinear force-free field model) is approximately 43°. Most of the accelerated electrons could be reflected at the footpoint region. This feature can be interpreted as brightenings produced by bouncing motion of the accelerated electrons.

*Unified Astronomy Thesaurus concepts:* Solar flares (1496); Solar radio emission (1522)

## 1. Introduction

Solar flares are thought to be caused by magnetic reconnection (Carmichael 1964; Sturrock 1966; Hirayama 1974; Kopp & Pneuman 1976). The released magnetic energy is converted into energy for particle acceleration and other processes. Although the mechanism of flare-associated particle acceleration remains controversial, researchers have proposed several models. Karlický & Kosugi (2004) argued that when a magnetic loop is contracting, the electrons supplemented by the loop are accelerated by an electric field perpendicular to the magnetic field lines. Another model proposes that electrons supplemented by a contracting loop are accelerated in a Fermi-like manner with bouncing motions along the magnetic field lines (Drake et al. 2006). Other models claim that the downflow collides with the loop top to create a "magnetic bottle" and is accelerated near the loop top under the turbulent electric field that mainly supports shock acceleration (Chen et al. 2015).

The above models are based on theoretical and observational knowledge of the pitch-angle distributions of accelerated electrons, which have been investigated through simulations and modeling approaches (Fleishman & Melnikov 2003; Simões & Costa 2006). For instance, Yokoyama et al. (2002) analyzed the high-speed propagation of a microwave source along a loop during a flare observed with the Nobeyama Radioheliograph (NoRH) on August 28 of 1999. By geometrically determining the length of the magnetic loop, they estimated the velocity of the accelerated electrons in the direction of the loop ($v_{\text{parallel}}$). Applying an approximation formula based on Bastian (1999) and Dulk (1985), Yokoyama et al. (2002) concluded that the actual speed $v$ of microwave-emitting electrons approximates light speed. They estimated a large pitch angle $\theta$ of the electrons emitting the propagating microwave source ($\sim$AA°). In other words, the nonthermal electrons are injected into the loop at high pitch angles. Thus far, Yokoyama et al.'s (2002) approach is the most straightforward one. Through one-dimensional simulations of electron motion along the loop and the time evolution of microwave emission, Minoshima & Yokoyama (2008) later demonstrated the same microwave-source propagation of incident electrons injected with an isotropic pitch-angle distribution.

The NoRH observes individual solar flares with a temporal resolution of 0.1 s, the timescale on which quasi-relativistic electrons travel along the loop. Although NoRH detected many flares after Yokoyama et al.'s (2002) event, no similar events have been reported. After carefully searching similar events, we found a suitable flare that suggests motion of accelerated electrons along a flare loop (an M8.7-class flare occurring on 2014 October 22). Section 2 of the present paper provides the observations of this flare and Section 3 analyzes the results. Section 4 interprets our observational results and discusses alternative possibilities.

## 2. Instruments

The NoRH radio interferometer, which observes the full disk of the Sun at 17 and 34 GHz, has successfully detected around 900 flare events between July of 1992 and March of 2020. The temporal resolution is usually 1 s. During a flare event, the NoRH collects the 0.1 s data in event mode. The spatial resolutions are approximately 15″ and 7″ at 17 and 34 GHz, respectively (Nakajima et al. 1994).

The Solar Dynamics Observatory (SDO) was launched by NASA in 2010 (Pesnell et al. 2011). The Atmospheric Imaging Assembly (AIA) (Lemen et al. 2011) on board the SDO can perform multiwavelength observations and obtain the relationship between the solar surface and atmospheric activity, solar wind, solar flares, and magnetic fields. We employed the AIA, which captures the structures of coronal loops with high-temperature ($\sim 10^7$ K) plasmas heated by a solar flare. The information of the coronal magnetic field was obtained from the nonlinear force-free field (NLFFF) database of the Institute of Space-Earth Environmental Research (Kusano et al. 2021). This database holds the three-dimensional magnetic field of the solar active regions analyzed by Kusano et al. (2020). The NLFFF was extrapolated from the photospheric vector magnetic field observed by the SDO/Helioseismic and Magnetic Imager (HMI) SHARP series (Bobra et al. 2014) using the magnetohydrodynamic relaxation method (Inoue et al. 2013).

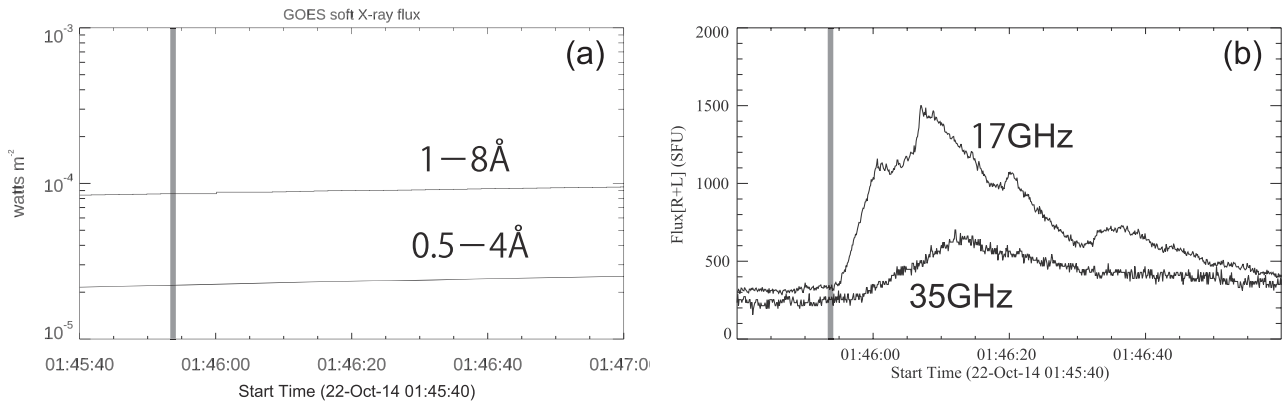

**Figure 1.** Temporal evolution of soft x-ray and microwave fluxes. (a) GOES soft X-ray flux (1–8 Å and 0.5–4 Å). (b) microwave flux observed with the Nobeyama radio polarimeter (17 and 35 GHz). The fast microwave propagation explained in the Section 3.1 was observed around the peak at 01:46 UT. The gray vertical lines in (a) and (b) indicate the location of $t = 0$ shown in Figure 3.

The Fermi Gamma-ray Burst Monitor (GBM) detects hard X-ray and gamma-ray bursts in the 8 keV to 40 MeV energy range (Atwood et al. 2009; Meegan et al. 2009). The temporal resolution of detecting flares or gamma-ray bursts is 0.064 s (Meegan et al. 2009). By analyzing the GBM data of hard X-rays, we discuss precipitations of accelerated electrons in the chromosphere.

The Reuven Ramaty High-Energy Solar Spectroscopic Imager (RHESSI) detected X-ray and gamma-ray images and spectra of solar flares in the energy range from 3 keV to 17 MeV (Lin et al. 2003). We discuss the location of precipitations of accelerated electrons analyzing the RHESSI data.

## 3. Results

The analyzed flare event occurred on October 22 of 2014 in NOAA Active Region (AR) 12192. The Geostationary Operational Environmental Satellite (GOES) X-ray class is M8.7. According to the NoRH event list,[3] the start time was 01:17:41 UT, the end time was 03:25:43 UT, and the peak time was 01:39:12 UT.

The flare event is summarized in Figure 1. Figure 1(a) shows the light curve of the soft X-ray flux observed with GOES during the observation time period discussed in this paper. Figure 1(b) shows the light curves observed at 17 and 35 GHz by the Nobeyama radio polarimeters (Nakajima et al. 1985). A microwave light curve of this flare displays several spikes. The most intense spike around 01:39:30 UT, not shown in Figure 1(b), is followed by a weaker spike around 01:46 UT. The present paper focuses on the second spike. Figure 2(a) overlays the 17 GHz NoRH image at 01:45:57.914 UT on the

---

magnetic field map observed with SDO/HMI at 01:34:15 UT (Figure 2(b)). Both ends of the brightest microwave loop structure locate in the eastern-negative/western-positive magnetic field region. To better understand the loop structure in the flaring region, we show the AIA 94 Å image in Figure 2(c). The loop containing the brightest (saturated) region identifies the microwave loop. The magnetic field information of this loop was numerically derived from NLFFF data using the HMI/SHARP data as boundary conditions. Figure 2(d) shows the calculated coronal loop based on the NLFFF data. Panels (a), (b), (c), and (e) of Figure 2 are helioprojective images observed from the earth and mapped in helioprojective Cartesian coordinates (HPC). In Figure 2(d), the images are remapped in cylindrical equal area (CEA) coordinates. The influence of these different coordinate systems is neglected because the flare appears near the disk center (see also Section 4.4). Figure 2(e) shows the hard X-ray and microwave sources and EUV flare ribbons. We can see that the western footpoints have different locations for hard X-ray and microwave. The hard X-ray and microwave footpoints are located at $(x,\ y) = (-310,\ -310)$ and $(-280,\ -325)$, respectively. Also, the bright point on the western flare ribbon extended to the south with time. Figure 2(f) is an overview of a loop emitting microwaves and hard X-rays at the footpoint, considered based on Figures 2(a), (c), and (e).

### 3.1. Cyclical Brightenings at Footpoint Regions

We determine the Points A and D as the eastern and western end of the main microwave source. In Figure 2 (c), AIA 94 Å image shows a lot of loops rooting at the Point A and one of them connects to the Point D. This means that the Points A and D are magnetically connected. This is also supported by the magnetic field reconstructed using NLFFF model. In Figure 2(d), there exist magnetic loops connecting between the Points A and D. Figure 3(b) shows a time variation of microwave intensity along the slit (magnetic loop connecting between the Points A and D) shown in Figure 3(a). Point B might be the starting point of the fast microwave propagation discussed in the Section 4.1. The loop top (Point C) is defined as the highest point on the magnetic field lines obtained from the NLFFF model. To eliminate the effects of beam size and image jitter in the image reconstruction, the data along the slit in Figures 3(b) was averaged over an area ($14''.7 \times 14''.7$) approximating the beam size. In addition, to minimize the

---

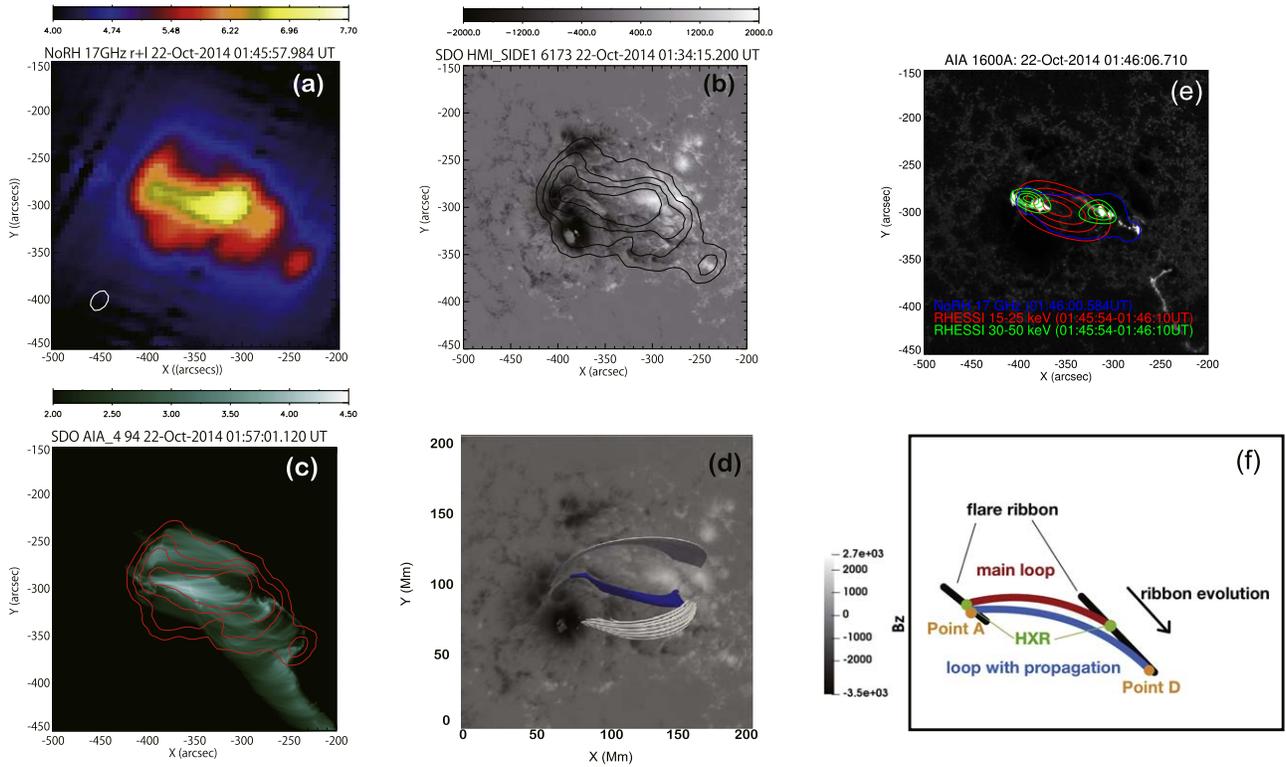

**Figure 2.** (a) 17 GHz image observed with NoRH at 01:45:57.984 UT. Color shows the logarithmic scale brightness temperature. The white contour at the lower left represents HPBW (half power beamwidth) shown as an indicator of the beam size at 17 GHz. (b) SDO/HMI data at 01:34:15 UT. The same image of (a) is shown as black contours. The contour levels are $10^5$, $10^{5.5}$, $10^6$, and $10^{6.5}$ K. (c) SDO/AIA (94 Å) logarithmic scale data at 01:57:120 UT. The same image of (a) is shown as red contours. The contour levels are $10^5$, $10^{5.5}$, $10^6$, and $10^{6.5}$ K. The loop with a saturated (brightest) area is well overlapped with the bright microwave loop in panel (a). (d) The blue and white lines represent the coronal magnetic field extrapolated by NLFFF approximation. The blue lines correspond to the saturated loops in (c) and the white lines show the other loops. The gray scale represents vertical magnetic field of SDO/HMI SHARP series at 01:36:00 UT. (e) AIA 1600 Å image at 01:46:06.710 UT. The red and green contours are RHESSHI data of 15–25 and 30–50 keV respectively. The blue contour is NoRH 17 GHz data. (f) Punch diagram of microwave loop and hard X-ray loop. The green and orange regions represent the footpoints of the loop radiating hard X-rays and microwaves, respectively. The size of panels (a), (b), and (c), is the same ($300'' \times 300''$). We cannot overlay an NoRH brightness temperature map on panel (d) because the coordinate system of panel (d) is different from that of panels (a), (b), and (c). $1''$ is about 720 km, so panel (d) is roughly the same size as the other panels ($277'' \times 277''$).

possibility of jitter, we employed data anchored within a substantial framework ($10^5$ K). We believe that averaging over a magnitude equivalent to that of the beam size serves to mitigate the possibility of jitter. Figure 3(b) shows the time variation of the brightness temperature at each point along the slit. In Figures 3(b), 4, and 5, $t = 0$ s corresponds to 01:45:54.084 UT. The magnetic field strength and length of the magnetic loops were computed using the NLFFF model (Figure 2(d)). The actual distances $L_{A–B}$, $L_{B–C}$, $L_{C–D}$, and $L_{A–D}$ between the points along the magnetic loop were 24.1, 40.4, 29.7, and 94.2 Mm, respectively. As shown in Figure 3(b), the microwave intensity increased twice at Point A. Figure 4 shows the time variations in brightness temperature and the brightness temperature difference (BTD) at Points A and D. The BTD is calculated as

$$\mathrm{BTD}(t, x) = f(t, x) - \sum_{k=0}^{30} \frac{f(t - 1.5 + 0.1k, x)}{31} \quad (1)$$

where $f(t, x)$ is the time variation of the brightness temperature along the slit and the BTD is the time average of the brightness temperature over 3.1 s at each points along the slit. As seen in the right panels of Figure 4, cyclical brightenings were observed at both footpoints (Points A and D). The peak times were 4.0 s ($t_{1A}$) and 6.1 s ($t_{2A}$) at Point A. As for Point D, the peak times were 4.6 s ($t_{1D}$) and 6.7 s ($t_{2D}$). It is likely that the injection of accelerated electrons occurred suddenly somewhere in the coronal loop. Using this observational feature, the velocity (Section 3.11–3.12) and pitch angle (Section 3.2) of the accelerated electrons were estimated in the two cases.

Note that the cyclical brightenings can be seen even when we created the microwave images with the other image-synthesis algorithm from the standard algorithm in Solarsoft that we used in this study, such as a self-calibration method developed by Dr. Stephen White.

### 3.1.1. Injection toward Both Footpoints

One interpretation of the cyclical brightenings is that two independent injection happens with an interval 2.1 s. In this case, the injection point should be located near the Point A rather than the loop top because the brightening at Point A is observed at the earlier time than that of Point D. In addition, the injection point should be the same for the two injections because the two time-lags, $t_{2A} - t_{1A}$ and $t_{2D} - t_{1D}$, are the same (0.6 s). The injection point is unknown. However, Point B is one possibility which is the starting point of a fast propagation of weak bright microwave source discussed in Section 4.1. Here, assuming the injection took place at Point B, the traveling times from Point B to Points A and D are written as $t_{B–A} = L_{A–B}/v$ and $t_{B–D} = L_{B–D}/v$, respectively, where $v$ is the propagation velocity along the loop. Letting $t_{B-D} - t_{B-A} = 0.6$ s, we obtain $v \sim 7.7 \times 10^4$ km s$^{-1}$. The time lag between the first and second injection is 2.1 s. This time lag is very similar to the bounce period along the loop (between Points A and D) with this velocity. Is this a coincidence? Thus, we consider another possibility in Section 3.1.2.

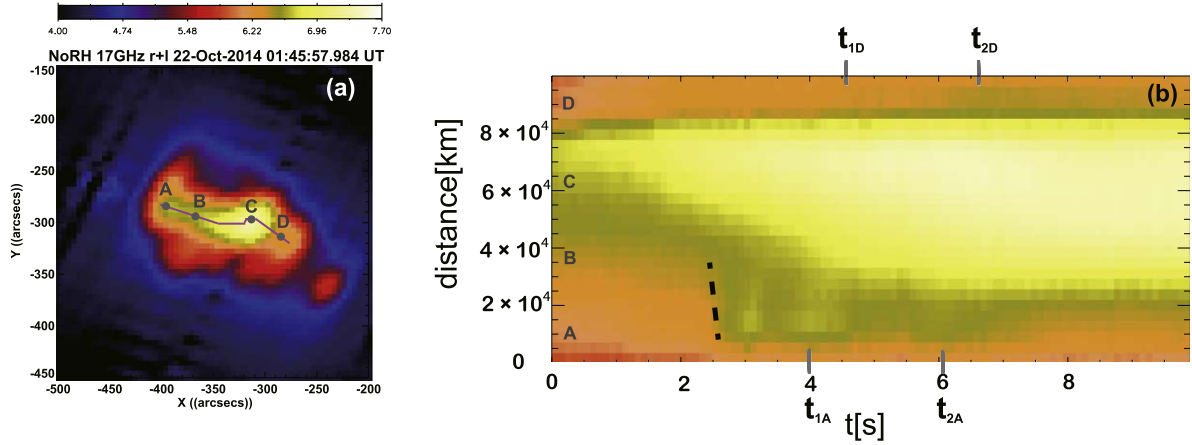

**Figure 3.** The analysis results of NoRH at 17 GHz (event mode). (a) The purple line denotes the slit along the apparent microwave propagation. (b) The time variation of the brightness temperature at each point along the slit. The base time of (b) is 01:45:54.084 UT ($t = 0$). A dotted line might show a high-speed microwave propagation from Point A to Point B discussed in Section 4.1. $t_{1A}$, $t_{2A}$, $t_{1D}$, and $t_{2D}$ are discussed in Section 3.1.

### 3.1.2. Propagation Velocity Assuming Bounce Motion

The second possibility is that the observed microwave-intensity enhancement at Point A can be caused by bouncing of the accelerated electrons between the footpoints. The peak times were 4.0 s ($t_{1A}$) and 6.1 s ($t_{2A}$) at Point A. Here we quantitatively discuss the possibility of bouncing motions at Point A. We suppose that the microwave source emitted by the accelerated electrons reached the corresponding point at these peak times ($t_{1A}$ and $t_{2A}$ at Point A). Assuming bouncing motion, the velocity of the high-energy electrons at Point A was estimated as $v_A = 2 \times L_{A-D}/\Delta t_A = 9.0 \times 10^4$ km s$^{-1}$, where $\Delta t_A = 2.1$ s represents the time intervals between the two peaks at Point A. As well as Point A, we assume that same accelerated electrons reached twice at Point D. The arrival time were 4.6 s ($t_{1D}$) and 6.7 s ($t_{2D}$). Moreover, the velocity of the high-energy electrons at Point D was estimated as $v_D = 2 \times L_{A-D}/\Delta t_D = 9.0 \times 10^4$ km s$^{-1}$, where $\Delta t_D = 2.1$ s represents the time intervals between the two peaks at Point D.

### 3.2. Estimating the Pitch Angle of Accelerated Electrons

In Bastian (1999), when the power-law index of accelerated electrons $\delta = 4$, the energy range of electrons contributing most to the 17 GHz gyrosynchrotron radiation is shown. Applying Dulk's approximation (Dulk 1985), we estimated $\delta$ using the $\alpha$ index obtained from the intensity ratio of NoRH 17 GHz and 34 GHz data. Figure 5 shows the map of the $\alpha$ (spectral index of microwave) and $\delta$ (spectral power-law index of the electrons). The $\delta$ index has various values along the loop, with values of 3–4 at $t = 4.0$ near Points A, B, and C along the loop. Compared to before the bounce period ($t = 2.4$), we can see that the $\delta$ index has temporarily softened along the loop. This suggests that there are many propagating electrons in the energy band mainly radiating at 17 GHz along the loop and that the $\delta$ index of the electrons is 3–4. From this fact, the $\delta$ index is determined to be 3–4 even for this flare, so we applied Bastian's approximation to this flare also. Bastian (1999) obtained the energy range of the electrons contributing to the 17 GHz radiation over the magnetic field range 200–1000 G. At 1000 G, the 17 GHz radiation was contributed mainly by electrons around 216 keV. In contrast, the magnetic field strengths $B_A$, $B_B$, $B_C$, and $B_D$ (where each subscript denotes a point) were 1400, 752, 654, and 1323 G, respectively. Bastian's approximation is inapplicable to Points A and D because the field strength near the footpoints exceeded 1000 G. Based on the results of Krucker et al. (2020), we suggested that the energy range of the electrons dominating the 17 GHz radiation is insensitive to magnetic field strengths above 1000 G. To simplify the situation, we assumed 1000 G at the footpoints although the actual magnetic field strength was higher at these points. At 1000 G, the energy of the accelerated electrons was 216 keV, so the electron speed was approximately $0.7c$. At 600 G, the energy and speed of the electrons were 462 keV and $0.85c$, respectively. For the first situation (Section 3.1.1), using $v = 7.7 \times 10^4$ km s$^{-1}$ as the apparent propagation speed along the loop, the pitch angle was roughly estimated as 69°–72° (where 69° and 72° correspond to 1000 and 600 G, respectively). As for the second case (Section 3.1.2), using $v = 9.0 \times 10^4$ km s$^{-1}$ as the apparent propagation speed along the loop, the pitch angle was roughly estimated as 78°–80° (where 78° and 80° correspond to 1000 and 600 G, respectively). If the accelerated electrons injected into the loop were traveling along the loop, the size of the loss cone at Point A can be determined from the magnetic-field-strength ratio at Points A and C. Assuming conservation of the first adiabatic invariant, the size of the loss cone was computed as $\theta_{loss} = \arcsin\sqrt{1/R}$, where $R = B_{max}/B_{min}$ is the mirror ratio ($B_{max}$ and $B_{min}$ denote the minimum and maximum field strengths along the magnetic loop, respectively). Substituting $B_{max} = B_A$ and $B_{min} = B_C$, $\theta_{loss}$ was roughly estimated as 43°. Because the estimated pitch angle of the electrons (69°–80°) in both cases is much larger than the loss cone angle, most of the electrons should have been reflected at Point A. The cyclical brightenings might show the bouncing motion of the accelerated electrons comparing the estimated pitch angle with the size of the loss cone. We cannot completely dismiss the possibility of jitter, though the cyclical pattern of the microwave increase at the footpoints could have suggested the bouncing motion of the accelerated electrons.

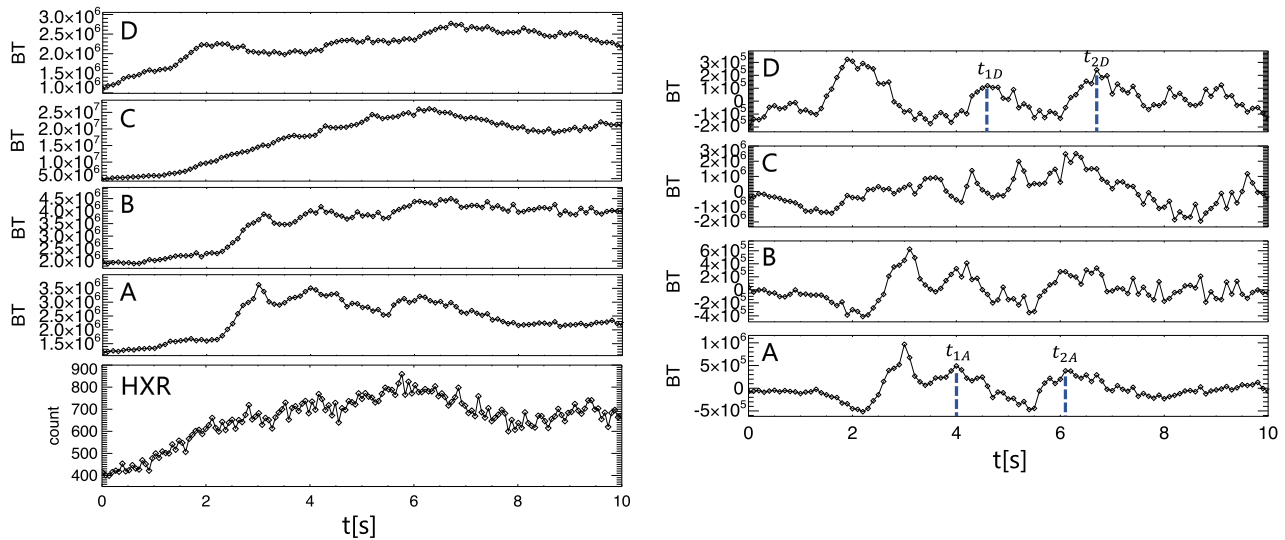

**Figure 4.** For the all panels in Figure 4, the time starts at 01:45:54.084 UT. The left five panels show the time variation of brightness temperature and the time variation of hard X-ray photon counts detected by the Fermi/GBM Sun-directed detector in the energy range from 49.0 to 101.4 keV. The right four panels show the BTD in Equation (1) from the Point A to Point D.

The bottom panel in Figure 4 shows the hard X-ray data observed with Fermi GBM. This plot provides a reference of electron precipitation in the chromosphere. The hard X-ray emissions were relatively intense during the period of bouncing motion and decreased after peaking at $t = 6.0$ s (the cessation time of bouncing motion). We interpreted that most of the accelerated electrons fell into the footpoints, for reasons that are currently unknown. Of course, as electrons emitting microwaves are more energetic than hard X-rays, there is no one-to-one correspondence between microwave propagation and the precipitation guessed from hard X-ray emissions. Observations confirm the presence of various loops along the flare ribbon, and a comparison with the RHESSHI image shows that the hard X-rays are emitted at the footpoint of a different loop than Point D, which is different from the loop we are discussing bouncing as a bounce period. Therefore, it is difficult to compare the time variation of microwaves and hard X-rays in the same time period. We can conclude only that the hard X-rays were enhanced during the microwave brightenings.

## 4. Discussions

### 4.1. Fast Propagation of Nonthermal Microwave Source

Figure 3 might show the fast microwave propagation observed along the loop as well as Yokoyama et al. (2002). Microwave propagation seems to start at Point B and appears near the footpoints (Points A and D) of the loop. The microwave source (green in Figure 3(b)) appears to propagate from Point B to Point A over an approximate distance of 22.4 Mm. From the travel time of the microwave source through this distance (0.3 s; see Figure 3(b)), the apparent propagation velocity was approximately $7.5 \times 10^4$ km s$^{-1}$ (dotted black line in Figure 3(b)). At $t = 2.4$ s, the microwave source was located at $(x, y) = (-365, -300)$ and extended in the northeast direction. The propagation velocity of the microwave source along the loop from Point B to Point A ($v_{B-A}$) was estimated as $8.1 \times 10^4$ km s$^{-1}$ rather than $7.5 \times 10^4$ km s$^{-1}$. This velocity is almost consistent with the velocity calculated in Section 3. This fact strength the assumptions and discussion in Section 3. Yokoyama et al.

(2002) observed the microwave fast propagation and discussed the pitch angle of accelerated electrons. In this research, we observed another microwave fast propagation and discuss the pitch angle using newer observations and simulation (SDO, RHESSI, Fermi, and NLFFF model). In Sections 3.1 and 3.2, we estimated the speed of microwave propagation, the pitch angle of accelerated electrons, and the size of loss cone using NLFFF model, which is not used in Yokoyama et al. (2002). Note that we had originally synthesized microwave images by using the standard algorithm, so-called "Hanaoka program" in the Solarsoft library, and the fast microwave propagation became unclear when we analyzing the data with the self-calibration method.

### 4.2. Microwave-intensity Variation in the Loop-top Region

In Figures 3(b) and 4, we have noted that the cyclical brightenings of microwaves at the footpoints may be a sign of bouncing motion of accelerated electrons. The region near Point C does not show that tendency because the brightness temperature of Point C in Figure 3(b) is 1 order of magnitude higher than at the footpoints. This bright loop-top region overwhelmed the faint microwave enhancement originating from bounce motions of the accelerated electrons. The brightness temperature at Point C remains higher than footpoints after the bouncing time. The brightness enhancement near the loop top (implying higher electron acceleration than near the footpoints) suggests the accumulation of accelerated electrons (Melnikov et al. 2002; Karlický & Kosugi 2004; Krucker et al. 2010). We do not discuss this topic here because it is beyond the scope of our paper. Moreover, as shown in Figures 2(e) and (f), the regions near the loop top are overlapped by various loops. As a result, this fact contributes to the brightness temperature becoming high.

### 4.3. Conditions for Detecting the Bouncing Motion of Accelerated Electrons

Bounce motion is thought to occur in other time of this flare and other flares. However, since electrons with various directions of motion exist simultaneously in the loop, it is difficult to capture the identical motion of specific electrons. In addition, the nonthermal microwave radiation depends on the viewing angle, which also depends on the location of the flare loops. The phenomena on the scale slower and smaller than the instrument's resolution cannot be captured directly. In order to observe electron bouncing motion, a flare event must meet the following three minimum conditions. (1) Electrons are injected into the flare loop with almost a delta function in time, resulting in microwave emission by bouncing electrons that is bright enough to be distinguished from background microwave emission. (2) The temporal and spatial resolution of the instrument is high enough to capture the loop size of the flare and the microwave fast propagation as shown in

$$\delta_{\mathrm{time}} \ll \frac{L_{\mathrm{loop}}}{v_{\mathrm{e}}} \qquad (2)$$

$$\delta_{\mathrm{space}} \ll L_{\mathrm{loop}} \qquad (3)$$

where $\delta_{\mathrm{time}}$, $\delta_{\mathrm{space}}$, $L_{\mathrm{loop}}$, and $v_{\mathrm{e}}$ are temporal resolution, spatial resolution, the length of the flare loop, and the speed of electrons (almost speed of light), respectively. (3) The injection of electrons occurs in the energy band of electrons that are easily visible at the instrument's observation frequency.

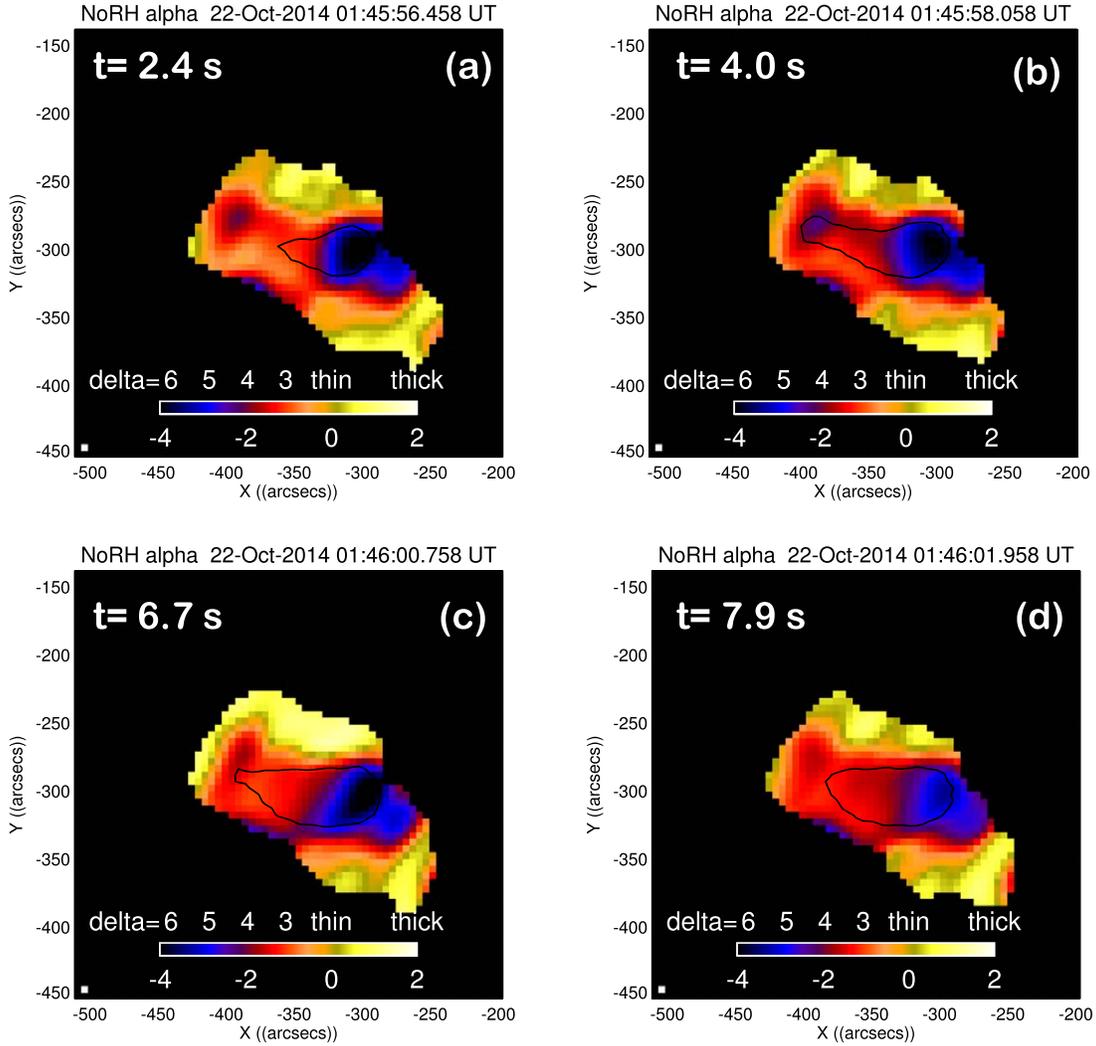

**Figure 5.** The $\alpha$ and $\delta$ indices map images calculated by the NoRH 17 and 34 GHz. Contour level of the NoRH 17 GHz for brightness temperature is $10^{6.5}$ K.

In this flare, we were able to observe the bouncing motion because the conditions described above were met in the case of the NoRH 17 GHz observation ($\delta_{\mathrm{time}} = 0.1$ and $\delta_{\mathrm{space}} = 15''$). While we performed the same analysis at 34 GHz as at 17 GHz, no fast propagation could not be identified at 34 GHz. Since the energy of the electrons contributing to the radiation are different between 17 and 34 GHz (Bastian et al. 1998), it might happen in this event that the propagating electrons have just suitable energy for the 17 GHz radiation rather than that for 34 GHz.

It is possible that even with delta electron injection, hard X-ray and microwave radiation may not result in sharp spikes for the following reasons. Regarding microwave radiation, The light curve could be the integration of the microwave radiation from the electrons already present in the loop and that from the microwave enhancement of the newly injected electrons. Furthermore, the light curve is not expected to be a delta-shaped spike due to the different timing of arrival at the footpoint due to pitch-angle scattering caused by Coulomb collisions. The energy band of electrons seen in hard X-ray radiation is lower in energy than that of electrons emitting microwaves, so the effect of pitch-angle scattering is likely to be even greater. An additional factor in this flare is that the hard

X-ray radiation is from the footpoints of a different loop than the one emitting the microwaves.

### 4.4. Validity of the NLFFF Model

The NLFFF model and NoRH images adopt different coordinate systems (CEA and HPC, respectively). We assumed negligible positional difference caused by the different coordinate systems because the flare event was close to the disk center. If a flare event occurs far from the disk center, the NoRH images and NLFFF results are not directly comparable. In such cases, the NLFFF model must adopt the spherical coordinate system to accurately generate the coronal magnetic field, and the projection effect must be considered properly.

The NLFFF model used in this study assumes static equilibrium. The magnetic field obtained by a data-driven magnetohydrodynamic simulation (Jiang et al. 2016; Guo et al. 2019; Kaneko et al. 2021), which does not assume dynamical equilibrium, might be more suitable for flare studies and should be investigated as a next step.

## 5. Conclusion

We detected cyclical brightenings at the footpoints during a solar flare occurring on October 22 of 2014. The high time resolution (0.1 s) of NoRH enables us to discuss the motion of the accelerated electrons moving along the loop. By analyzing the NoRH data, we found that the bright microwave feature propagated from a loop leg (Point B) to a footpoint (Point A). Using the coronal magnetic field and NLFFF models, we approximated the propagation speed of the bright microwave feature along the loop as $7.7 \times 10^4 \mathrm{~km~s^{-1}}$ and $9.0 \times 10^4$. The pitch angle of the high-speed electrons propagating through the magnetic loop was approximately $69°$–$80°$. Considering the coronal magnetic field derived by the NLFFF model, the size of the loss cone was around $43°$ suggesting that most of the accelerated electrons are reflected by a magnetic mirror near the footpoints. This is the first flare event that discusses accelerated electrons through cyclical brightenings by comparing observations (microwave, EUV, hard X-ray) and the NLFFF model. Before the cyclical brightenings at the footpoints, a fast microwave propagation might be observed from Point B to Point A. The propagation velocity was approximated as $8.1 \times 10^4 \mathrm{~km~s^{-1}}$, consistent with the propagation velocity derived from the cyclical brightenings ($7.7 \times 10^4$ and $9.0 \times 10^4 \mathrm{~km~s^{-1}}$). We obtained suggestions for possible magnetic mirror reflection and accelerated electron bouncing between footpoints. The time variation of the brightness temperature at the footpoints reinforced the bounce motion of the accelerated electrons along the loop. The accuracy of estimating physical quantities such as loop length, coronal magnetic field strength, and loss cone size has greatly improved since Yokoyama et al.'s (2002) era, owing to SDO/AIA observations and progress in NLFFF modeling, although some errors remain. Questions regarding this event also remain. For example, why are the nonthermal electrons seemingly injected at only one footpoint? Injections of high-energy electrons at Point B might be related to loop-to-loop interactions, as is often observed with NoRH (Hanaoka 1999). One-side injection might be related to asymmetries in the electron-acceleration process or the magnetic field structure. To answer this question, we must compare our observations with those of computer-simulated electron motions and microwave emissions under particular boundary conditions.

Gyrosynchrotron radiation is complicated, so we need to simulate this flare comparing our observations. Actually, Minoshima & Yokoyama (2008) claimed that fast microwave propagation is blinded to electron motion when the initial pitch-angle distribution is narrow. In future work, we plan to model the fast propagation of microwaves by combining the NLFFF model with simulations. For verification, the pitch-angle distributions will be compared with the simulation results.

### Acknowledgments

We thank K. Watanabe, M. Shimojo, N. Narukage, and K. Iwai for their helpful comments, K. Kusano for use of the ISEE NLFFF database, and S. Krucker and S. M. White for the discussion and analysis of RHESSHI and NoRH. This work is partly supported by JSPS KAKENHI, grant No. JP18H01253 and JP23K03455. T.K. was supported by the National Center for Atmospheric Research, a major facility sponsored by the National Science Foundation under Cooperative Agreement No. 1852977. We would like to express our sincere gratitude for the financial support provided by ISEE/CICR (Institute for Space - Earth Environmental Research/Center for International Collaborative Research) for overseas travel.

### ORCID iDs

Keitarou Matsumoto  https://orcid.org/0000-0003-2002-0247
Satoshi Masuda  https://orcid.org/0000-0001-5037-9758
Takafumi Kaneko  https://orcid.org/0000-0002-7800-9262

### References

Atwood, W. B., Abdo, A. A., Ackermann, M., et al. 2009, ApJ, 697, 1071
Bastian, T. S., Benz, A. O., & Gary, D. E. 1998, ARA&A, 36, 131
Bastian, T. S. 1999, in Impulsive Flares: A Microwave Perspective, Proc. of the Nobeyama Symp., ed. T. S. Bastian, N. Gopalswamy, & K. Shibasaki , 211
Bobra, M. G., Sun, X., Hoeksema, J. T., et al. 2014, SoPh, 289, 3549
Carmichael, H. 1964, in The Physics of Solar Flares, Proc. of the AAS-NASA Symp., 50, ed. W. N. Hess (Washington, DC: National Aeronautics and Space Administration), 451
Chen, B., Bastian, T. S., Shen, C., et al. 2015, Sci, 350, 1238
Drake, J. F., Swisdak, M., Che, H., & Shay, M. A. 2006, Natur, 443, 553
Dulk, G. A. 1985, ARA&A, 23, 169
Fleishman, G. D., & Melnikov, V. F. 2003, ApJ, 587, 823
Guo, Y., Xia, C., Keppens, R., Ding, M. D., & Chen, P. F. 2019, ApJL, 870, L21
Hanaoka, Y. 1999, PASJ, 51, 483
Hirayama, T. 1974, SoPh, 34, 323
Inoue, S., Magara, T., Pandey, V. S., et al. 2013, ApJ, 780, 101
Jiang, C., Wu, S. T., Feng, X., & Hu, Q. 2016, NatCo, 7, 11522
Kaneko, T., Park, S.-H., & Kusano, K. 2021, ApJ, 909, 155
Karlický, M., & Kosugi, T. 2004, A&A, 419, 1159
Kopp, R. A., & Pneuman, G. W. 1976, SoPh, 50, 85
Krucker, S., Hudson, H. S., Glesener, L., et al. 2010, ApJ, 714, 1108
Krucker, S., Masuda, S., & White, S. M. 2020, ApJ, 894, 158
Kusano, K., Iijima, H., Kaneko, T., et al. 2021, ISEE Database for Nonlinear Force-Free Field of Solar Active Regions, doi:10.34515/DATA.HSC-00000
Kusano, K., Iju, T., Bamba, Y., & Inoue, S. 2020, Sci, 369, 587
Lemen, J. R., Title, A. M., Akin, D. J., et al. 2011, in The Solar Dynamics Observatory, ed. P. Chamberlin, W. D. Pesnell, & B. Thompson (New York: Springer), 17
Lin, R. P., Dennis, B. R., & Benz, A. O. 2003, The Reuven Ramaty High-Energy Solar Spectroscopic Imager (RHESSI) Mission Description and Early Results (Dordrecht: Springer)
Meegan, C., Lichti, G., Bhat, P. N., et al. 2009, ApJ, 702, 791

Melnikov, V. F., Shibasaki, K., & Reznikova, V. E. 2002, ApJL, 580, L185
Minoshima, T., & Yokoyama, T. 2008, ApJ, 686, 701
Nakajima, H., Kawashima, S., Shinohara, N., et al. 1994, IEEEP, 82, 705
Nakajima, H., Sekiguchi, H., Sawa, M., Kai, K., & Kawashima, S. 1985, PASJ, 37, 163

Pesnell, W. D., Thompson, B. J., & Chamberlin, P. C. 2011, The Solar Dynamics Observatory (New York: Springer), 3
Simões, P. J. A., & Costa, J. E. R. 2006, A&A, 453, 729
Sturrock, P. A. 1966, Natur, 211, 695
Yokoyama, T., Nakajima, H., Shibasaki, K., Melnikov, V. F., & Stepanov, A. V. 2002, ApJL, 576, L87